\documentclass{ws-procs11x85}
\usepackage{multicol}
\usepackage{amsfonts}
\usepackage{amsthm}
\usepackage{amssymb}
\usepackage[mathscr]{euscript}
\usepackage{t1enc}
\usepackage{amsmath}
\usepackage{amscd}
\usepackage{psfrag}
\usepackage{graphics}
\begin{document}
\title{How to Quantize Forces\,(?):\\
${}$\\
An Academic Essay How the Strings Could Enter Classical Mechanics}
\author{Denis Kochan}
\address{Department of Theoretical Physics and Physics Education\\
FMFI UK, Mlynsk\' a dolina F2, 842 48 Bratislava, Slovakia\\
$\&$\\
Quniverse, L\' i\v s\v cie \' udolie 116, 841 04 Bratislava, Slovakia\\
$\&$\\
Department of Theoretical Physics\\
Nuclear Physics Institute AS CR, 250 68 \v{R}e\v{z}, Czech Republic\\
\vspace{0.3cm}
\texttt{kochan@fmph.uniba.sk}}
\date{}
\maketitle
\abstract{Geometrical formulation of classical mechanics with forces that are not necessarily
potential-generated is presented. It is shown that a natural geometrical ``\,playground\,'' for a mechanical system
of point particles lacking Lagrangian and/or Hamiltonian description is an odd dimensional line element contact bundle.
Time evolution is governed by certain canonical two-form $\Omega$ (an analog of $dp\wedge dq-dH\wedge dt$), which
is constructed purely from forces and the metric tensor entering the kinetic energy of the system.
Attempt to ``\,dissipative quantization\,'' in terms of the two-form $\Omega$ is proposed. The Feynman's path integral
over histories of the system is rearranged to a ``\,world-sheet\,'' functional integral. The ``\,umbilical
string\,'' surfaces entering the theory connect the classical trajectory of the system and the
given Feynman history. In the special case of potential-generated forces, ``\,world-sheet\,'' approach precisely
reduces to the standard quantum mechanics. However, a transition probability amplitude expressed in terms of
``\,string functional integral\,'' is applicable (at least academically) when a general dissipative environment
is discussed.\\



\centerline{\emph{Dedicated to Karla and Mario Ziman's on the occasion of their wedding, and to one sunny smiling friend.}}}

%\begin{multicols}{1}
\baselineskip=13.07pt

\section{Introduction}

Classical mechanics is the best elaborated and understood part of physics. At the first sight relatively innocent
Newton's equation of motion becomes mathematically very interesting and fruitful when one leaves a flat vector space,
for example, by imposing a simple set of constraints. Nowadays geometrical description of mechanics is concentrated
around beautiful and powerful mathematical artillery, which includes\cite{arnold}${}^-$\cite{lewis} symplectic
and/or Poisson geometry, contact structures, jet prolongations, Riemannian geometry, variational calculus, ergodic
theory and so on. Advantages and disadvantages of any of these approaches are dependent on intended applications and
personal preferences and disposals. Some of them are useful, when one goes from classical to quantal. Another, when
one wants to pass from the non-relativistic domain to relativistic one. And the third and fourth, when we attempt to
generalize discrete system dynamics to the continuum one and/or when the number of particles is so large that some
statistical methods should be imposed.

The aim of the paper is to provide a geometrical picture of classical mechanics for physical systems
for which Lagrangian and/or Hamiltonian description is missing. This means that the forces acting within the system
are not potential-generated. After explaining the geometry beyond the classical dissipative dynamics we make
an attempt at quantization.
We avoid to couple the system to an environment and to form one conservative super-system switching on an interaction.
Our approach is based solely on the system under study and a dynamics in which dissipative strength effects of the
environment are described by a velocity dependent external force. In the case
of potential-generated forces, the proposed description becomes equivalent to the standard
canonical formalism.

The paper is organized as follows. The first two preliminary subparagraps deal with
some basic facts about the contact geometry and its application in point particle mechanics. We concentrate ourselves
on the definition of the line element contact bundle and its internal geometrical structure (smooth atlas, bundleness,
canonical distribution, natural lift of curves).
The subsequent introduction to mechanics is standard. We just want to remind the reader of basic notation and convince
him that to use an extended configuration space and related line element contact structure in mechanics is highly
advantageous.
The main attention is paid to a correct geometrical setting of forces, i.e. we are looking for the answer to the question:
``\,what type of tensorial quantities are forces in general?\,'' The guiding object of (dissipative) dynamics
which governs the time evolution is certain two-form $\Omega$. It is constructed only from forces and kinetic energy.
The last paragraph is a rather academic and speculative elaboration on possible quantization in terms of Feynman
functional integral. ``\,Umbilical string\,'' surfaces naturally enter the quantization and transition amplitudes are
obtained by a ``\,world-sheet\,'' functional integration.

To be honest and collegial, it is necessary to provide here some standard references on the quantization
of dissipative systems. There exist several different approaches\cite{kanai}${}^-$\cite{dodonov}
(explicitly time-dependent Hamiltonian, method of dual coordinates, nonlinear Schr\"{o}dinger equation, method of the
loss-reservoir) and an interested reader could try to focus on one of the following keywords: damped
(phil)harmonic oscillator, Kostin's nonlinear Schr\"{o}dinger-Langevin equation, Caldirola-Kanai equation,
stochastic quantization, Caldeira-Leggett model.

\section{Preliminaries: Line Element Contact Bundle and Classical Mechanics}

By a mechanical system we understand throughout the paper a system of particles whose positions and velocities
are restricted by a set of \emph{holonomic} and/or \emph{integrable differential} constraints.
The constraints as well as exterior forces (which are not supposed to be potential-generated only) can be
explicitly time dependent.\\
The aim of the following subsections is to remind the reader of the geometrical setting that is necessary
for the proper description of the time evolution. Hopefully, we will recover soon that a natural ``\,playground\,'' for
classical mechanics is the \emph{line element contact bundle} of an extended configuration space.

\subsection{Line Element Contact Bundle}

A beautiful introduction to contact structures in physics with a variety of applications can be found
in the William Burke's book\cite{burke}. I am very strongly recommending to go through it in details.
Its eloquent motto: \emph{...how in hell you can vary $\dot{q}$ without changing $q$...} applies also to the
following text.

Let $\mathscr{M}$ be an ordinary $(n+1)$-dimensional smooth real manifold and
$\gamma_1,\,\gamma_2:\mathbb{R}\rightarrow\mathscr{M}$ two parameterized curves
thereon. One says that $\gamma_1$ and $\gamma_2$ are in
\emph{contact} at a common point $p\in\gamma_1\cap\gamma_2\subset\mathscr{M}$, if their tangent vectors (instant velocities)
at that point are proportional to each other. The contactness is obviously a weaker notion than
tangentiality.

A \emph{line contact element} at point $p$ is the equivalence class of curves being in contact at $p$. Practically,
to give a line contact element means to choose a point $p$ of $\mathscr{M}$ and to fix a one-dimensional subspace
(undirected line) $\ell\subset T_p\,\mathscr{M}$. The set of all undirected lines passing through the origin of the
tangent space under consideration is the projective space $\mathbb{P}\,(T_p\,\mathscr{M})$.
Thus forming a sum:
$$
\mathscr{CM}:=\bigcup\limits_{p\in\mathscr{M}}\mathbb{P}\,(T_p\,\mathscr{M})\equiv(\mathbb{P}\,T)\,\mathscr{M}
$$
we get a set of all line contact elements of the manifold $\mathscr{M}$. $\mathscr{CM}$ is not a structureless object,
it inherits smooth structure from $\mathscr{M}$ that turns it into a manifold. Concisely, let
$\{\mathcal{O}_\mathfrak{I},\,\varphi_\mathfrak{I}\}_\mathfrak{I}$ be any smooth atlas of $\mathscr{M}$ and
$\{T(\mathcal{O}_\mathfrak{I}),\,\Phi_\mathfrak{I}\}_\mathfrak{I}$ induced local trivialization of
its tangent bundle $T\,\mathscr{M}$, i.e.
$$
\Phi_\mathfrak{I}: T(\mathcal{O}_\mathfrak{I})\longrightarrow\varphi_\mathfrak{I}(\mathcal{O}_\mathfrak{I})\times\mathbb{R}^{n+1}\,, \ \ \
\bigl\{p\in\mathcal{O}_\mathfrak{I}\,,v\in T_p\,(\mathcal{O}_\mathfrak{I})\bigr\}\longmapsto\bigl(q^0,\dots,q^n\,\bigl|\,\dot{q}^{\,0}=v^0,\dots,\dot{q}^{\,n}=v^n\bigr)\,.
$$
Let us, moreover, define the system of:
\begin{itemize}
\item[-] open subsets $\mathscr{C}_a(\mathcal{O}_\mathfrak{I})\subset T(\mathcal{O}_\mathfrak{I})$  ($a$ runs from $0$ to $n$):
$$
\mathscr{C}_a(\mathcal{O}_\mathfrak{I}):= \Phi^{-1}\Bigl\{
\,\mbox{those points of}\ \varphi_\mathfrak{I}(\mathcal{O}_\mathfrak{I})\times\mathbb{R}^{n+1}\
\mbox{whose $a$-th dot coordinate $\dot{q}^{\,a}\neq 0$}\Bigr\}
$$
\item[-] morphisms $\Phi_{a,\,\mathfrak{I}}$ $:=$ the restriction $\Phi_\mathfrak{I}\bigr|_{\mathscr{C}_a(\mathcal{O}_\mathfrak{I})}$
\end{itemize}
Then the collection $\{\mathscr{C}_a(\mathcal{O}_\mathfrak{I}),\,\Phi_{a,\,\mathfrak{I}}\}_{a,\,\mathfrak{I}}$ provides a
smooth atlas of $\mathscr{CM}$.

Down to earth, the point $(p\,,\ell\,)\in\mathscr{CM}$ is the one-dimensional subspace $\ell\subset T_p\,\mathscr{M}$.
As such, it can be represented as a linear envelope of a vector $v\in T_p\,\mathscr{M}$ whose, let us say $a$-th,
coordinate w.r.t. $\{\partial_{q^0}\bigr|_p,\dots,\partial_{q^n}\bigr|_p\}$, is equal
to one, i.e.
$$
\ell=\bigl\{w\in T_p\,\mathscr{M}:\ w=\Bbbk v\,, \mbox{where}\ \Bbbk\in\mathbb{R}-\{0\}\ \mbox{and}
\ v=v^0\,\partial_{q^0}\bigl|_p+\cdots+1\,\partial_{q^a}\bigl|_p+\cdots+v^n\,\partial_{q^n}\bigl|_p\bigr\}=:[v]\,.
$$
Then $(p\,,\ell\,)$ belongs to the chart $\mathscr{C}_a(\mathcal{O}_\mathfrak{I})\subset\mathscr{CM}$
and
$$
\Phi_{a,\,\mathfrak{I}}\bigl((p,\,\ell\,)\bigr)=
\bigl(\,q^0(p)\,,\dots,\,q^n(p)\,\bigl|\,\dot{q}^{\,0}=v^0\,,\dots,\,\dot{q}^{\,a-1}=v^{a-1}\,,\,\dot{q}^{\,a+1}=v^{a+1}\,,\dots,\,\dot{q}^{\,n}=v^n\,\bigr)\in\varphi_\mathfrak{I}(\mathcal{O}_\mathfrak{I})\times\mathbb{R}^{n}\,.
$$
One can write down transition functions $(\Phi_{a,\,\mathfrak{I}})\circ(\Phi_{a^\prime,\,\mathfrak{I}^\prime})^{-1}$
over the non-empty overlaps
$\mathscr{C}_a(\mathcal{O}_\mathfrak{I})\bigcap\mathscr{C}_{a^\prime}(\mathcal{O}_{\mathfrak{I}^\prime})$ in the explicit
way and verify their smoothness and compatibility on the triple intersections. It is, in fact, not necessary, since
everything follows from appropriate modifications of the smooth consistent atlas of the tangent bundle $T\,\mathscr{M}$.

Conclusion: $\mathscr{CM}$ is a $(2n+1)$-dimensional smooth manifold which, moreover, forms an $n$-dimensional bundle
over $\mathscr{M}$. Clearly, when sending the line contact element $(p\,,\ell\,)\in\mathscr{CM}$ to its contact point
$p\in\mathscr{M}$, we get the smooth bundle map $\tau:\,\mathscr{CM}\rightarrow\mathscr{M}$. Note that the fiber
$\tau^{-1}(p)$ is compact space $\mathbb{P}\,(T_p\,\mathscr{M})$. This bundle is called the
\emph{line element contact bundle} $\mathscr{CM}$.

Apart from a closed set of measure zero, any line contact element $\ell$ at a point $p\in\mathscr{M}$ can be represented
by a specially chosen contact curve $\gamma_{\mathrm{spec}}$. Down to earth, let
$\gamma: s\mapsto q^{0}=q^{0}(s),\dots,\,q^n=q^n(s)$ be any curve such that $\gamma(s_o)=p$, $\tfrac{d}{dt}\,q^0(s_o)\neq 0$ and
$[\tfrac{d}{dt}\,\gamma(s_o)]=\ell$. Then when restricting ourselves to a sufficiently small neighborhood of the point $p$,
we can reexpress initial parameter $s=f(q^0)$ as a function of the local coordinate $q^{0}$. An advantageous
representative of $(p\,,\ell\,)$ can be then provided by the equivalence class of the curve
$\gamma_{\mathrm{spec}}: q^0\mapsto q^{0}=q^0,\,q^1=q^1(f(q^0)),\dots,\,q^n=q^n(f(q^0))$.
Therefore further, when it will become computationally necessary,\footnote{It will be especially useful in
mechanics where $\mathscr{M}$ corresponds to an extended configuration space $\mathbb{R}\times\mathcal{Q}$ and all
physically relevant trajectories are of that form.} we will break down the natural equivalency of the local charts in the
atlas of $\mathscr{CM}$ under consideration and prefer the subsystem
$\{\mathscr{C}_0(\mathcal{O}_\mathfrak{I}),\Phi_{0,\,\mathfrak{I}}\}_\mathfrak{I}$, i.e. the local
coordinate basis $(t:=q^0\,|\,q^1,\dots,\,q^n\,|\,\dot{q}^{\,1},\dots,\,\dot{q}^{\,n})$.

$\mathscr{CM}$ itself has an additional internal structure. Apart from being a bundle $\tau:\mathscr{CM}\rightarrow\mathscr{M}$
it admits a canonical $(n+1)$-dimensional distribution $\mathfrak{C}\subset T\,(\mathscr{CM})$. For a given point
$(p\,,\ell\,)$, a subspace $\mathfrak{C}_{(p,\ell)}\subset T_{(p,\,\ell)}\,\mathscr{CM}$ is specified as follows:
$$
\mathfrak{C}_{(p,\,\ell)}:=\bigl\{w\in T_{(p,\,\ell)}\,\mathscr{CM},\ \mbox{such that}\ \tau_*(w)\in\ell\subset T_p\,\mathscr{M}\bigr\}=\bigl(\tau_*\bigr)^{-1}(\ell)\,.
$$
If $(p\,,\ell\,)\in\mathscr{C}_0(\mathcal{O}_\mathfrak{I})$, then there is a vectorial ``\,precursor\,''
$v=\partial_t|_p+v^i\,\partial_{q^i}|_p\in T_p\,\mathscr{M}$ such that $\ell=[v]$. Here and further in this
subparagraph, index $i$ runs from $1$ to $n$.
Tangent vector $w$ at the line element contact bundle point $(p\,,\ell\,)$ and its $\tau$ push-forward at $p$
are expressed as follows:
$$
w=M\,\partial_t\Bigl|_{(p,\,\ell)}\,+\,N^i\,\partial_{q^i}\Bigl|_{(p,\,\ell)}\,+\,O^i\,\partial_{\dot{q}^{\,i}}\Bigl|_{(p,\,\ell)}\,,  \ \ \ \ \ \ \ \
\tau_*(w)=M\,\partial_t\Bigl|_{p}\,+\,N^i\,\partial_{q^i}\Bigl|_{p}\,.
$$
Here the numbers $M\neq 0$, $N^i$, $O^i$ stand for $(2n+1)$ components of $w$ and $\tau_*(w)$ w.r.t. our special operative
coordinate basis.

In order that $\tau_*(w)\in\ell$, the $n$-tuple of coefficients $N^i$ should be equal to $Mv^i$.
The remaining $(n+1)$ components are ``\,free of commission,'' and therefore $\dim(\mathfrak{C}_{(p,\,\ell)})=(n+1)$
as was stated.
\begin{figure}[htp]
\begin{center}
\psfrag{t}{$\tau$}
\psfrag{p}{$p$}
\psfrag{l}{$\ell$}
\psfrag{h}{$\ell^\prime$}
\psfrag{tp}{$\tau^{-1}(p)$}
\psfrag{pl}{$(p\,,\ell\,)$}
\psfrag{ph}{$(p\,,\ell^\prime\,)$}
\psfrag{Cpl}{$\mathfrak{C}_{(p,\,\ell)}$}
\psfrag{Cph}{$\mathfrak{C}_{(p,\,\ell^\prime)}$}
\psfrag{w}{$w$}
\psfrag{tw}{$\tau_*(w)$}
\psfrag{M}{$\mathscr{M}$}
\psfrag{CM}{$\mathscr{CM}$}
\psfrag{TpM}{$T_p\,\mathscr{M}$}
\psfrag{TplCM}{$T_{(p,\,\ell)}\,\mathscr{CM}$}
\psfrag{TphCM}{$T_{(p,\,\ell^\prime)}\,\mathscr{CM}$}
\epsfxsize=7cm
\epsfbox{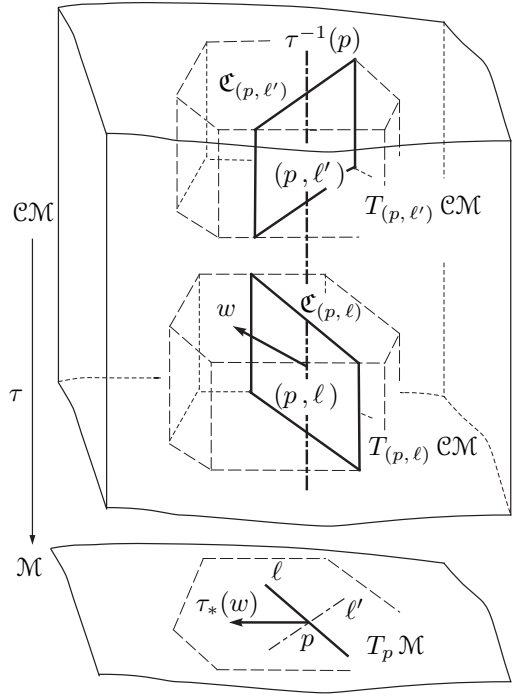}
\caption{Schematic picture of the canonical distribution $\mathfrak{C}$ at two $\mathscr{CM}$ points $(p\,,\ell\,)$ and
$(p\,,\ell^\prime\,)$ at the same fiber $\tau^{-1}(p)$.}\label{picture}
\end{center}
\end{figure}

Practically, in order to describe the above canonical distribution $\mathfrak{C}$, one can use any Pfaff's system
of $n$ algebraical one-forms. This means that for each point $(p\,,\ell\,)\in\mathscr{CM}$ there is a collection
of co-vectors $\alpha^i\in T^*_{(p,\,\ell)}\,\mathscr{CM}$, each of which annihilates
the canonical subspace $\mathfrak{C}_{(p,\,\ell)}$. There is no reason to assume that $\alpha$'s are varying smoothly
with the point of $\mathscr{CM}$; the Pfaff's system is algebraical and not differential.
Any regular linear combinations of $\alpha$'s define an equivalent system of
annihilators of $\mathfrak{C}_{(p,\,\ell)}$ at the point $(p\,,\ell\,)$. What one can try to do is to adjust (locally at
least) the linear combinations of the annihilating co-vectors in such a way that the new system would be smoothly depending on
$(p\,,\ell\,)$. There is obviously no canonical way how to do this; locally it is always possible, but
still ambiguous.\footnote{For example, one can impose some Euclidian metric on a local patch of $\mathscr{CM}$ and
then identify $T^*_{(p,\,\ell)}\,\mathscr{CM}$ and $T^*_{(p^\prime,\,\ell^\prime)}\,\mathscr{CM}$ in that
patch with the help of parallel transport.}\\
In our special coordinate system $\mathscr{C}_0(\mathcal{O}_\mathfrak{I})$ we can use for example the collection of the
following differential one-forms:
\begin{equation}\label{alpha-system}
\alpha^i=dq^i-\dot{q}^{\,i}dt\ \ \ \ \ \ \ \ i=1,\dots,n\,.
\end{equation}
The canonical $(n+1)$-dimensional distribution $\mathfrak{C}$ is not integrable, which follows immediately from the Frobenius
theorem ($\alpha^i\wedge d\alpha^i=dq^i\wedge dt\wedge d\dot{q}^{\,i}\neq 0$ for $\forall$ $i=1,\dots, n$).\footnote{
Regardful reader surely realized that the Einstein summation convention is used when the indices labeled by the
same letter are matching each other in the superscript and subscript positions only, otherwise, like for example in
$\alpha^i\wedge d\alpha^i$, the summation is not performed.}

Any locally smooth curve $\gamma: \mathbb{R}\rightarrow\mathscr{M}$, $s\mapsto\gamma(s)$ can be naturally lifted to
a line element contact bundle curve $\widehat{\gamma}: \mathbb{R}\rightarrow\mathscr{CM}$. By definition,
$\widehat{\gamma}$ assigns to the given value of the parameter $s$ the contact point $\gamma(s)$ and the undirected
line $\ell(s):=[\tfrac{d}{ds}\,\gamma(s)]$. After rewriting it in coordinates one realizes that the pull-back
$\widehat{\gamma}^{\,*}(\alpha^i)=0$ for all $i=1,\dots,n$.
The converse is also true: if there is a contact bundle curve $\widehat{\lambda}: \mathbb{R}\rightarrow\mathscr{CM}$ such that
$\widehat{\lambda}^*(\alpha^i)=0$, then $\widehat{\lambda}$ is the lift of some base curve $\lambda$. Thus we have a
simple criterion enabling us to recognize which line element contact bundle curve is originally coming from ``\,down-stairs\,''
and which is the ``\,native resident\,'' of $\mathscr{CM}$.

\subsection{Classical Mechanics on $\mathscr{CM}$}

Let us start with the Newton-Lagrange philosophy. From an observer point of view, a mechanical system with $n$ degrees of
freedom occupies at a given instant of time (external parameter defined by the ticking of the observer's watch)
a point in a certain configuration space. Geometrically it is a $n$-dimensional smooth manifold. In applications,
it mostly emerges after imposing certain number of constraints on some background flat space $\mathbb{R}^{3N}$.
The constraints are supposed to be holonomic and explicitly time dependent (it is possible to consider
in a very similar fashion also time dependent integrable differential constraints). Mathematically they are given by
a set of algebraical equations in $\mathbb{R}^{3N}$ containing time $t$ as an external parameter.
For each time $t$ there ``\,survives\,'' some (sub)manifold $\mathcal{Q}^t\subset\mathbb{R}^{3N}$ whose points satisfy
the whole system of constraint equations.
The explicit time dependence is easier to handle if one passes to an extended space
by adopting observer's time $t$ as a new coordinate and visualizing the ``\,surviving\,'' sets in a single
space-time picture as $\Lambda:=\bigcup_{t}\,(t,\,\mathcal{Q}^{t})\subset\mathbb{R}[t]\times\mathbb{R}^{3N}$.
The Lagrange novelty was to introduce a parametric manifold of generalized coordinates $\mathcal{Q}$
and map it by some one-parameter family\footnote{To be really ultrarigorous, the one-parameter family here and the one
defined above by the observer's watch are two different mathematical sets. In general they should be connected by some
complicated one-to-one mapping $f$. It is a nice habit to chose $f$ as identity, i.e. physically we are
using a couple of synchronized watches to measure time in the extended parametric space $\mathbb{R}[t]\times\mathcal{Q}$
and in the physical space-time $\mathbb{R}[t]\times\mathbb{R}^{3N}$.} of diffeomorphisms $\{\varphi^t\}_t$ in such a way that
$\varphi^t(\mathcal{Q})=\mathcal{Q}^t$. Equivalently, one can form an \emph{extended parametric space}
$\mathbb{R}[t]\times\mathcal{Q}$ and define the single diffeomorphism $\Phi:\mathbb{R}[t]\times\mathcal{Q}\rightarrow\Lambda,$
$(t,\,\mathcal{Q})\mapsto(t,\varphi^t(\mathcal{Q}))$.

\begin{figure}[tbh]
\begin{center}
\psfrag{QRt}{$\mathbb{R}[t]\times\mathcal{Q}$}
\psfrag{Phi}{$\Phi$}
\psfrag{t}{$t$}
\psfrag{Q}{$\mathcal{Q}$}
\psfrag{Rt}{$\mathbb{R}[t]$}
\psfrag{Qt}{$\mathcal{Q}^t$}
\psfrag{L}{$\Lambda\subset\mathbb{R}[t]\times\mathbb{R}^{3N}$}
\epsfxsize=12cm
\epsfbox{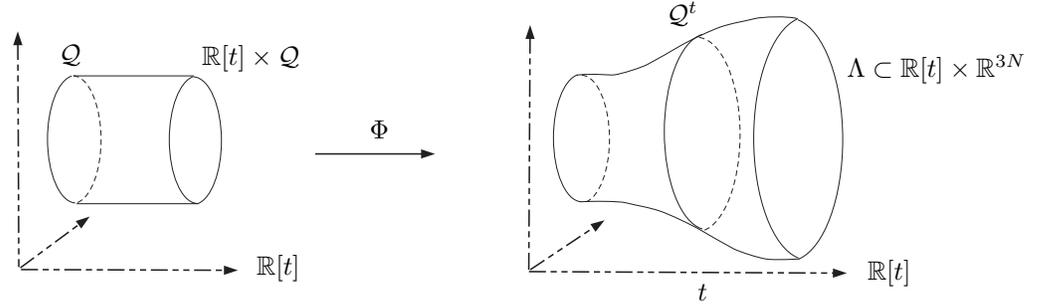}
\hspace{-3cm}
\caption{Embedding of the extended parametric space into the observer's space-time realized by $\Phi$.}\label{obraz}
\end{center}
\end{figure}

This was just kinematics; to be able to describe dynamics one should know where relevant physical data are hidden.
Within the ambient space-time $\mathbb{R}[t]\times\mathbb{R}^{3N}$ all the necessary physical information is given in
the following geometrical objects:
\begin{itemize}
\item[-] \underline{the kinetic energy $G$}:
\begin{itemize}
\item[$\circ$] originally, $G$ is constant Riemannian metric, i.e. co-variant symmetric rank two-tensor field on
$\mathbb{R}^{3N}$, which in practical application takes the standard form:
$$
G=\sum\limits_{k=1}^{N}\frac{1}{2}\,m_k\bigl\{dx^k\otimes dx^k+dy^k\otimes dy^k+dz^k\otimes dz^k\bigr\}\,,
$$
here $(x^1,y^1,z^1,x^2,\dots,x^N,y^N,z^N)$ are global cartesian coordinates on $\mathbb{R}^{3N}$,
\item[$\circ$] it describes the ``\,inertia\,'' properties of the physical matter,
\item[$\circ$] for any point $p\in\mathbb{R}^{3N}$, tensor $G$ gives us the quadratic form on
$T_p\,\mathbb{R}^{3N}\ni v\mapsto G_p\,(v,v)\in\mathbb{R}$, therefore from a global point of view, $G$ is a quadratic
function at fibers over the tangent bundle $\pi:T\,\mathbb{R}^{3N}\rightarrow\mathbb{R}^{3N},$ $(p\,,v)\mapsto p$
(we are not introducing a special letter to distinguish $G$ as function and as a tensor),
\item[$\circ$] $G$ can be pulled-back from the tangent bundle to the extended tangent bundle
$\mathbb{R}[t]\times T\,\mathbb{R}^{3N}$ w.r.t. the obvious second factor projection
$\mathbb{R}[t]\times T\,\mathbb{R}^{3N}\rightarrow T\,\mathbb{R}^{3N}$ (we
are proposing that the constituent weights $m_k$ are constant),
\end{itemize}
\item[-] \underline{the acting forces $Q$}:
\begin{itemize}
\item[$\circ$] at each instant of time, $Q^{t}$ is a horizontal one-form over the tangent bundle
$\pi:T\,\mathbb{R}^{3N}\rightarrow\mathbb{R}^{3N}$, i.e. for any point
$(p\,,v)\in T\,\mathbb{R}^{3N}$ and any vertical vector $w\in T_{(p,v)}(T\,\mathbb{R}^{3N})$ ($\pi_*(w)=0$)
it holds $w\,\lrcorner\,Q^{t}=0$,
\item[$\circ$] setting $Q^{t}$ at time $t$ to be the tangent bundle one-form reflects the fact that the forces
are in general not only $p=\,$position but also $v=\,$velocity dependent,
\item[$\circ$] a horizontal co-vector $Q^{t}\in T^*_{(p,v)}\,(T\,\mathbb{R}^{3N})$ might be turned into
a co-vector $\widetilde{Q}_{v}^{t}\in T^*_p\,(\mathbb{R}^{3N})$;
concisely, let $u$ be a tangent vector to the background space $\mathbb{R}^{3N}$ at a point $p=\pi(p\,,v)$ and
$u^{\uparrow}$ its arbitrarily performed lift to $T_{(p,v)}\,(T\,\mathbb{R}^{3N})$ such that $\pi_*(u^{\uparrow})=u$,
then when setting $u\,\lrcorner\,\widetilde{Q}_{v}^{t}:=u^{\uparrow}\,\lrcorner\,Q^{t}$ we get a well defined
co-vectorial object,\footnote{Horizontality
of $Q$ ensures the independence on lift procedure (in fact two different lifts $u^\uparrow$ and $u^{\Uparrow}$
which project by $\pi$ onto the same base vector $u$, differ by a vertical vector); since all other operations leading
to $\widetilde{Q}_v^{t}$ are linear, $\widetilde{Q}_v^{t}$ is a linear functional on $T_p\,(\mathbb{R}^{3N})$.}
\item[$\circ$] the physically measurable effect of the force $Q^{t}$ is this: if the system occupies at some
instant of time $t$ a configuration $(p\,,v)$, the elementary work $\delta W$ of $Q^{t}$ at an infinitesimal (virtual)
displacement $\delta r\in T_p\,(\mathbb{R}^{3N})$ is given by $\delta W=\widetilde{Q}_{v}^{t}(\delta r)$,
\item[$\circ$] fixing the coordinates $(x^1,y^1,z^1,\dots|\dot{x}^{\,1},\dot{y}^{\,1},\dot{z}^{\,1},\dots)$ in the tangent bundle
$T\,\mathbb{R}^{3N}$ we get the expression $Q^{t}:=Q_{x^k}(t)\,dx^k+Q_{y^k}(t)\,dy^k+Q_{z^k}(t)\,dz^k$; keep in mind that $3N$
components are position, velocity, as well as explicitly time dependent functions,
\item[$\circ$] for better handling of the explicit time dependence of the forces, it is more handy to convert $Q^{t}$ at time $t$
to the horizontal one-form over a sub-bundle $i_{t}:T\,\mathbb{R}^{3N}\stackrel{\sim}{\rightarrow}\{t\}\times T\,\mathbb{R}^{3N}\hookrightarrow\mathbb{R}[t]\times T\,\mathbb{R}^{3N}$;
thus in global we get a horizontal (also called a \emph{semi-basic}) differential one-form $Q$ over
$\mathbb{R}[t]\times T\,\mathbb{R}^{3N}$ with the property $Q^{t}=i_t^*(Q)$ (do not overlook that
$\partial_t\,\lrcorner\,Q=0$)
\end{itemize}
\end{itemize}
A \emph{space of all physical states}\footnote{It is one of the postulates of mechanics that having chosen at some time $t$
position(s) $p$, velocity(ies) $v$ and knowing the acting physical forces afterwords, it is possible to predict the history when solving the dynamical
equations.} of the extended parametric space is the manifold\footnote{Some authors
call the extended parametric space $0$-\emph{jet} $J^{0}\bigl(\mathbb{R},\,\mathcal{Q}\bigr)$ and its state
space $1$-\emph{jet} $J^{1}\bigl(\mathbb{R},\,\mathcal{Q}\bigr)$.} $\mathbb{R}[t]\times T\,\mathcal{Q}$.
It contains not just the time $t$ and the generalized coordinates $(q^1,\dots,q^n)$ that cover some patch of the parametric
space $\mathcal{Q}$, but also the generalized velocities $(\dot{q}^{\,1},\dots,\dot{q}^{\,n})$. All dynamical data
are given after we perform the pull-back of $G$ and $Q$ from $\mathbb{R}[t]\times T\,\mathbb{R}^{3N}$ to the state
space $\mathbb{R}[t]\times T\,\mathcal{Q}$ using the diffeomorphism $\Phi$ and/or its
differential $d\Phi$.
What follows is a well known story. One needs to solve the Lagrange's equations which determine generalized
accelerations in terms of generalized positions, velocities and forces, as well as time:
\begin{equation}\label{Lagrange eq.}
\frac{d}{dt}\left(\frac{\partial\,\mathbb{T}}{\partial\, \dot{q}^{\,i}}\right)
-\frac{\partial\,\mathbb{T}}{\partial\, q^{i}}=(\mathbb{Q}\,)_i\ \ \ \ \
i=1,\dots, n\,.
\end{equation}
Here a bit sloppy but short notation is introduced. $\mathbb{T}:=\Phi^*G$ and $\mathbb{Q}:=\Phi^*Q$
represent the pull-backs of $G$ and $Q$ w.r.t $\Phi$ and/or $d\Phi$.\\
Expressing from (\ref{Lagrange eq.}) the accelerations $\tfrac{d}{dt}\,\dot{q}^{\,i}\equiv\ddot{q}^{\,i}$ as functions\footnote{It
is implicitly supposed that the matrix of the second partial derivatives $\partial_{\dot{q}^{\,i}}\partial_{\dot{q}^{\,j}}(\,\mathbb{T}\,)$ is
invertible.}
$f^i(q,\,\dot{q},\,\mathbb{Q},\,t)$, the integration of the Lagrange's equations of motion becomes equivalent to
the problem of finding the integral curves of the vector field
\begin{equation}\label{Lagrange fld.}
\dot{\gamma}=\partial_{t}\,+\,\dot{q}^{\,i}\,\partial_{q^i}\,+\,f^i(q,\,\dot{q},\,\mathbb{Q},\,t)\,\partial_{\dot{q}^{i}}
\end{equation}
over the space of all physical states $\mathbb{R}[t]\times T\,\mathcal{Q}$. After determining the integral curve $\gamma$
that satisfies at time $t$ the given initial conditions, we can project it onto the extended parametric space curve
$\gamma_{_{\mathbb{R}\times\mathcal{Q}}}$ forgetting about its velocity. To see how the
motion looks like in the ``\,real space-time\,'' we finally map $\gamma_{_{\mathbb{R}\times\mathcal{Q}}}$ by the
diffeomorphism $\Phi$ onto the curve
$\Phi(\gamma_{_{\mathbb{R}\times\mathcal{Q}}})\subset\Lambda\subset\mathbb{R}[t]\times\mathbb{R}^{3N}$.\\

The extended parametric space $\mathbb{R}[t]\times\mathcal{Q}$ is what I called the extended configuration space
in the previous section. It forms the $(n+1)$-dimensional smooth manifold and for obvious reasons let us use
the symbol $\mathscr{M}$ for it instead of a bit impractical $\mathbb{R}[t]\times\mathcal{Q}$.
The time $t$ is a distinguished coordinate on $\mathscr{M}$. This enables us to identify
the $(2n+1)$-dimensional space of all the physical states $\mathbb{R}[t]\times T\,\mathcal{Q}$ with the
open dense set in the line element contact bundle $\mathscr{CM}$. Down to earth, the mapping
$$
\mathbb{R}[t]\,\times\,T\,\mathcal{Q}\,\ni\,(t\,|\,q^{\,1},\dots,\,q^{\,n}\,|\,v^{\,1},\dots,\,v^{\,n})
\ \longleftrightarrow\
p=(t\,,\,q^1,\dots,\,q^n)\in\mathscr{M}\ \ \mbox{and}
\ \ \ell=\bigl[\partial_{t}\bigl|_{p}+v^{\,i}\,\partial_{q^i}\bigl|_{p}\bigr]\subset T_{p}\mathscr{M}
$$
gives us the identification $\mathbb{R}[t]\times T\,\mathcal{Q}\,\simeq\,\mathscr{C}_0\mathscr{M}$ in the explicit form.
What is the advantage of proceeding in that way? Mainly the observation that $\mathscr{CM}$ supports the canonical
$(n+1)$-dimensional distribution $\mathfrak{C}$. Note that at any point $(p\,,\ell\,)\in\mathscr{C}_0\mathscr{M}$ the dynamical
vector:
$$
\dot{\gamma}\Bigl|_{(p,\ell)}=\partial_{t}\Bigl|_{(p,\ell)}\,+\,\dot{q}^{\,i}\,\partial_{q^i}\Bigl|_{(p,\ell)}\,+
\,f^i\bigl(q,\,\dot{q},\,\mathbb{Q}(q,\dot{q},t),\,t\bigr)\,\partial_{\dot{q}^{i}}\Bigl|_{(p,\ell)}\in T_{(p,\ell)}\,\mathscr{CM}
$$
defines a one-dimensional subspace $[\dot{\gamma}]_{(p,\ell)}\subset T_{(p,\ell)}\,\mathscr{CM}$. Therefore
there emerges certain one-dimensional distribution over (the open dense set of) $\mathscr{CM}$.
The first sight inspection of $\dot{\gamma}|_{(p,\ell)}$ immediately shows that $[\dot{\gamma}]_{(p,\ell)}$ belongs to the
canonical space $\mathfrak{C}_{(p,\,\ell)}\subset T_{(p,\ell)}\,\mathscr{CM}$.
To put it in other words, any curve $\gamma$ that integrates the vector field (\ref{Lagrange fld.}) satisfies
$\gamma^*(\alpha^i)=0$ over the Pfaff's system (\ref{alpha-system}), which determines the canonical distribution
$\mathfrak{C}$. As a consequence, there exists a unique base curve
$\gamma_{_{\mathbb{R}\times\mathcal{Q}}}:\mathbb{R}\rightarrow\mathscr{M}$ such that $\gamma=\widehat{\gamma}_{_{\mathbb{R}\times\mathcal{Q}}}$.

If one succeeds to find another system of $n$ one-forms over $\mathscr{CM}$, let us call them $\beta_i$, such that
$\beta$'s and $\alpha$'s would be linearly independent and $\beta$'s would annihilate the subspace $[\dot{\gamma}]_{(p,\ell)}$
for each $(p\,,\ell)\in\mathscr{CM}$, one would have a Pfaff's system of $2n$ one-forms, which will completely describe
$1$-dimensional distribution $[\dot{\gamma}]$ over the $(2n+1)$-dimensional line element contact bundle $\mathscr{CM}$.
It is not a big deal to verify that this is satisfied if we put
\begin{equation}\label{beta-system}
\beta_i:=d\bigl\{\partial_{\,\dot{q}^{i}}\,\mathbb{T}\bigr\}
-\bigl\{\partial_{\,q^i}\,\mathbb{T}\bigr\}dt
-(\mathbb{Q}\,)_i\,dt\ \ \ \ \ \ i=1,\dots,n\,.
\end{equation}\\

Let us stop for the moment and recapitulate what we have found out about the dynamics until now. We
have seen that the complete time evolution of a mechanical system subjected to known forces is given by the vector
field $\dot{\gamma}$. This follows from the dynamical postulate: the Lagrange's equations (\ref{Lagrange eq.}).
Now let us continue; the different perspective of classical dynamics is coming. Using the known function
$\mathbb{T}$ and the force one-form $\mathbb{Q}=(\mathbb{Q}\,)_i\,dq^i$ over the line element contact bundle, we can
establish the two-form $\Omega$:
\begin{equation}\label{Omega}
\Omega=\alpha^i\,\wedge\,\beta_i=-\mathbb{Q}\,\wedge\,dt-d\bigl\{\mathbb{T}\,dt+(\partial_{\,\dot{q}^{i}}\,\mathbb{T})\,\alpha^i\bigr\}\in\Gamma\Bigl(\bigwedge\hspace{-1mm}{}^2\,T^*\,\mathscr{CM}\Bigr)
\end{equation}
Inspection of (\ref{Omega}) shows that $\Omega$ is non-singular\footnote{A differential two-form $\Omega$ over an
odd dimensional manifold $M$ is \emph{non-singular}, if its null spaces are one-dimensional for any point of $M$.}
and its null-spaces are exactly the one-dimensional subspaces $[\dot{\gamma}]_{(p,\ell)}\subset T_{(p,\ell)}\,\mathscr{CM}$. From here, there is just
a little step to recover the full dynamics. Indeed, by picking up at each such null subspace $[\dot{\gamma}]_{(p,\ell)}$
a vector $v$ for which $v\,\lrcorner\,dt\bigl|_{(p,\ell)}=1$, we are point-wisely reconstructing the dynamical vector
field (\ref{Lagrange fld.}).

Isn't it wonderful? Let me explain why I am seeing it to be so interesting. Unforgettable explanation of the canonical
formalism of classical mechanics can be found in the chapter 9 of the excellent book\cite{arnold}
of Vladimir Arno\v{l}d. It is shown there that the vortex lines\footnote{... integral submanifolds of a $1$-dimensional
distribution given by the null spaces of $d\omega^1$ ...} of the one-form $\omega^1=p_i\,dq^i-H\,dt$
on the $(2n+1)$-dimensional extended phase space $\mathbb{R}[t]\times T^*\mathcal{Q}$ are just the integral curves
of the canonical equations of Hamilton. Few lines bellow you can find the footnoted sentences ...\emph{The form $\omega^1$ seems here to appear
out of thin air. In the following paragraph we will see how the idea of using this form arose from optics}... Ok, a small
difference is here, since we are occupying $\mathbb{R}[t]\times T\mathcal{Q}$ instead of the extended phase space,
but the unifying idea leading to the equations of mechanics is the same, namely, the null spaces. Here of the two-form
$\Omega$, there of $d\omega^1$, here we get dynamics in the
``\,Lagrange picture,'' there in the ``\,Hamilton one.'' The remarkable difference is that $\Omega$ is not a closed
two-form in general (and thus not locally exact), its cohomology class is specified by the two-form
$-\mathbb{Q}\,\wedge\,dt$ and therefore its reasonable optical analog is somehow missing (at least from my point of
view). There is a legitimate
question: Is there a $\mathscr{CM}$ function $\mathbb{U}$ such that
$-\mathbb{Q}\,\wedge\,dt=d\bigl\{\mathbb{U}\,dt+(\partial_{\,\dot{q}^{i}}\,\mathbb{U})\,\alpha^i\bigr\}$?
The answer is notorious from the basic course of analytical mechanics: the function $\mathbb{U}$ we ask for should
satisfy:
$$
(\mathbb{Q}\,)_i=-\partial_{\,q^i}\,\mathbb{U}+\frac{d}{dt}\,\Bigl(\partial_{\,\dot{q}^i}\,\mathbb{U}\Bigr)\,.
$$
In that happy case
$\Omega=-d\bigl\{(\mathbb{T}-\mathbb{U})\,dt+\partial_{\,\dot{q}^{i}}\,(\mathbb{T}-\mathbb{U})\,\alpha^i\bigr\}=:-d\theta_{\,\mathbb{L}}$,
and the Lagrange's function $\mathbb{L}:=\mathbb{T}-\mathbb{U}$ can be introduced.
Moreover, $\omega^1=p_i\,dq^i-H\,dt$ converts after an appropriate Legendre's transformation to the Cartan's one-form
$\theta_{\,\mathbb{L}}$. So having potential-generated forces everything looks like it should. There is
a one-form $\omega^1\,\leftrightsquigarrow\,\theta_{\,\mathbb{L}}$ and a safe way to its quantization. But how should
one proceed if the forces are such that the ``\,precursor\,'' $\theta_{\,\mathbb{L}}$ is missing and all what is
applicable is represented by $\Omega$? Let us postpone the investigation of that problem to the following paragraph.

\section{Quantization: Path \emph{vs.} Surface Integral }

In the previous section we have observed that the classical evolution can be completely described by finding the
integrating submanifolds of the distribution of null spaces of the two-form $\Omega$. So we could claim:
``\,classical mechanics is only $\Omega$-sensitive.'' Everything else is a bonus valid in special cases only.
We were being impractical not to use the (local) potential $\theta_{\,\mathbb{L}}$ or the Lagrangian $\mathbb{L}$ which
would enable us
to investigate the invariants and/or conserved quantities. But the physical principles are constituted
over the equations of motion, not over the Lagrangian or Hamiltonian themselves. This resembles an instant soup:
if you have got it just pour it in the water and that's it, but not every soup at all tastes like this one ...

On the other hand, it seems that quantum mechanics is rather $\theta_{\,\mathbb{L}}$ (or if you
wish $\omega^1$)-sensitive. You are surely familiar with all this gossip about optical-mechanical analogy, presenting
classical mechanics as some limit (i.e. just as approximation) of ``\,wave mechanics\,'' whose wave fronts are
specified by $\theta_{\,\mathbb{L}}$ and whose Huygens' principle is expressed by the Hamilton-Jacobi equation. I like
it very much, but most impressive way (at least from my point of view) how to relate classical and quantal lies in
the Feynman path-integral approach.

According to the Feynman prescription:\cite{feynman-hibbs}${}^,$\cite{faddeev-slavnov} the probability amplitude of the
transition of the system from the extended configuration space event $\mathrm{e}_{_{-}}:=(t^{\,{}}_{_-},\,q^{\,1}_{_-},\dots,\,q^{\,n}_{_-})$ to
$\mathrm{e}_{_{+}}:=(t^{\,{}}_{_+},\,q^{\,1}_{_+},\dots,\,q^{\,n}_{_+})$ is:
\begin{equation}\label{FeynmanI}
\mathbf{A}(\mathrm{e}_{_{-}},\,\mathrm{e}_{_{+}})\ \propto\,
\boldsymbol{\int} [\mathscr{D}\gamma]\,\exp{\Bigl\{\frac{i}{\hbar}\int\limits_{\gamma}\,\theta_{\,\mathbb{L}}\Bigr\}}\,.
\end{equation}
The ``\,path-summation\,'' here is taken over the set of all curves\footnote{Do not miss that all the curves
entering the ``\,path-summation\,'' are parameterized exclusively by the observer's time, i.e.
$\gamma:\,t\mapsto \bigl(t=t,\,q^i=q^i(t),\,\dot{q}^{\,i}=\dot{q}^{\,i}(t)\bigr)$.} $\gamma$ on
$\mathscr{CM}\xrightarrow{\tau}\mathscr{M}$ such that their $\tau$-projections satisfy
$\tau(\gamma)(t_{_-})=\mathrm{e}_{_{-}}$
and $\tau(\gamma)(t_{_+})=\mathrm{e}_{_{+}}$.
The exponent in (\ref{FeynmanI}) is the standard integral of the one-form $\theta_{\,\mathbb{L}}$ carried
out on the line element contact bundle curve $\gamma$. The question about the
``\,measure\,'' $[\mathscr{D}\gamma]$ and the proper normalization of the probability amplitude $\mathbf{A}$
are, fortunately, not a subject of our discussion. Let me remind the reader that the probability amplitude formula
(\ref{FeynmanI}) is used less frequently than its extended phase space version. When expressing generalized velocities
$\dot{q}^{\,i}$ in $\theta_{\,\mathbb{L}}$ in terms of generalized momenta $p_i=\tfrac{\partial\mathbb{L}}{\partial\dot{q}^{\,i}}$,
we get $\mathbf{A}$ in terms of the functional integral in the extended phase space $\mathbb{R}[t]\times T^*\mathcal{Q}$:
$$
\mathbf{A}(\mathrm{e}_{_{-}},\,\mathrm{e}_{_{+}})\ =\,
\boldsymbol{\int} [\mathscr{D}\widetilde{\gamma}]\,\exp{\Bigl\{\frac{i}{\hbar}\int\limits_{\widetilde{\gamma}}\,p_i\,dq^i-Hdt\Bigr\}}\,,
\ \ \mbox{where one can formally set}\ \
[\mathscr{D}\widetilde{\gamma}]=\dfrac{dp_{_+}}{2\pi}\,\prod\limits_{\overset{t\in(t_{_-},t_{_+})}{{}}}\dfrac{dp_{_t}\,dq_{_t}}{2\pi}\,.
$$
The two formulas for the amplitude $\mathbf{A}$ are equivalent; the bunch of curves $\gamma$ and
$\widetilde{\gamma}$ that enters the functional integrations are connected by the same type of Legendre's transformation
as the one-forms $\theta_{\,\mathbb{L}}$ and $\omega^1$.

The mentioned sensitiveness of quantum mechanics on the one-form $\theta_{\,\mathbb{L}}\leftrightsquigarrow\omega^1$ is
evident.
In what follows, we propose some modifications leading to the replacement of $\theta_{\,\mathbb{L}}$
by the two-form $\Omega$. These would enable us to ``\,quantize\,'' also dissipative forces. In the special case
when they are conservative ($\Omega=-d\theta_{\,\mathbb{L}}$) our prescription will be equivalent with Feynman's.

\begin{figure}[tbh]
\begin{center}
\psfrag{t}{$t$}
\psfrag{q}{$q$}
\psfrag{qd}{$\dot{q}$}
\psfrag{c}{$\gamma$}
\psfrag{cc}{$\gamma_{\mathrm{class}}$}
\psfrag{em}{$\mathrm{e}_{_-}$}
\psfrag{ep}{$\mathrm{e}_{_+}$}
\psfrag{tem}{$\tau^{-1}(\mathrm{e}_{_-})$}
\psfrag{tep}{$\tau^{-1}(\mathrm{e}_{_+})$}
\psfrag{tc}{$\tau(\gamma)$}
\psfrag{tcc}{$\tau(\gamma_{\mathrm{class}})$}
\psfrag{lm}{$\lambda_{_-}$}
\psfrag{lp}{$\lambda_{_+}$}
\psfrag{tm}{$t_{_-}$}
\psfrag{tp}{$t_{_+}$}
\psfrag{tau}{$\tau$}
\psfrag{CM}{$\mathscr{CM}$}
\psfrag{M}{$\mathscr{M}$}
\epsfxsize=7cm
\epsfbox{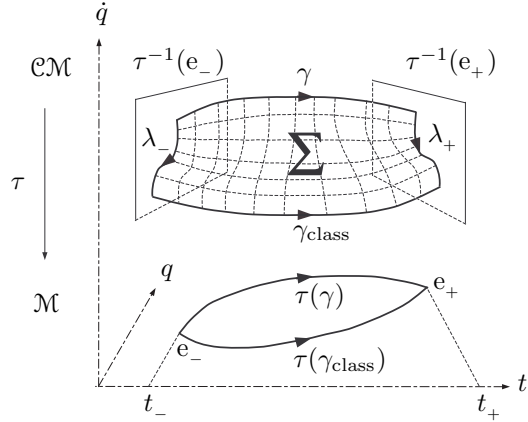}
\vspace{-1cm}
\caption{``\,Umbilical surface\,'' $\Sigma$ connects classical trajectory $\gamma_{\mathrm{class}}$ with the history
$\gamma$. Sideways boundary curves $\lambda_{_\pm}$ are located within the $n$-dimensional submanifolds
$\tau^{-1}(\mathrm{e}_{_\pm})$. On the figure these submanifolds are schematically represented just by the two-\-dimensional ``\,D-branes.\,''}
\label{figure}
\end{center}
\end{figure}

The class of curves entering the ``\,path-summation\,'' in (\ref{FeynmanI}) has one simple characteristic: initial and
final endpoint of any admissible curve $\gamma$ should belong to fiber submanifolds
$\tau^{-1}(\mathrm{e}_{_-})\subset\mathscr{CM}$ and $\tau^{-1}(\mathrm{e}_{_+})\subset\mathscr{CM}$.
In between this class, there is one special curve, the classical trajectory\footnote{We optimistically propose that
solutions of the equations of motion might be ``\,inverted\,'' on relatively broad interval of time, i.e. from given position
at the final time we are be able to adjust the velocity at the initial time in such a way that the system will evolve
uniquely into the prescribed endpoint.} $\gamma_{\mathrm{class}}$.
Using it, we get for any $\gamma$ within this class (oriented) 1-cycle:
$$
\partial\,\Sigma:=\gamma+\lambda_{_+}-\gamma_{\mathrm{class}}-\lambda_{_-}\,.
$$
Here $\lambda_{_-}$ and $\lambda_{_+}$ are arbitrarily chosen curves belonging to the fiber submanifolds $\tau^{-1}(\mathrm{e}_{_-})$
and $\tau^{-1}(\mathrm{e}_{_+})$ that join the initial and final points of $\gamma$ and $\gamma_{\mathrm{class}}$,
respectively (see Figure \ref{figure}).

Moreover, the restriction\footnote{When looking at (\ref{Omega}), one immediately realizes that the same is true for the
two-form $\Omega$.}
of the Cartan's one-form $\theta_{\,\mathbb{L}}$ to any fiber of $\tau:\mathscr{CM}\rightarrow\mathscr{M}$ is trivial,
therefore both integrals of $\theta_{\,\mathbb{L}}$ carried out over the $\lambda$'s vanish automatically and we can write:
\begin{equation}\label{I}
\int\limits_{\gamma}\,\theta_{\,\mathbb{L}}\,-\int\limits_{\gamma_{\mathrm{class}}}\hspace{-0.5mm}\theta_{\,\mathbb{L}}\,=\,
\int\limits_{\partial\,\Sigma}\,\theta_{\,\mathbb{L}}\,=\,\int\limits_{\Sigma}\,d\theta_{\,\mathbb{L}}\,,
\ \ \Sigma\ \mbox{is any $\mathscr{CM}$-surface whose boundary}\ \ \partial\,\Sigma=\gamma+\lambda_{_+}-\gamma_{\mathrm{class}}-\lambda_{_-}\,.
\end{equation}
Let me remind you that
\begin{itemize}
\item[-] the second term on the left hand side of (\ref{I}) is just the value of the classical action $S_{\mathrm{class}}$
\item[-] the existence of ``\,umbilical\,'' $\mathscr{CM}$-surface $\Sigma$ that connects given curve $\gamma$ with $\gamma_{\mathrm{class}}$
         is determined by topological properties\footnote{Topological properties we are talking about are ``\,measured\,''
         by the fundamental group $\Pi_1(\mathscr{M})$. For obvious reasons we are cowardly skipping off any
         discussion of the quantization in topologically nontrivial cases.} of
         $\mathscr{M}$, e.g. when $\mathscr{M}$ is simply-connected then any
         1-cycle $\partial\,\Sigma$ is at the same time a 1-boundary of some 2-chain $\Sigma$
\end{itemize}
Motivated by (\ref{I}) and encouraged by the Feynman's thesis\cite{feynman} sentence: ...\emph{the central mathematical
concept is the analogue of the action in classical mechanics. It is therefore applicable to mechanical systems whose
equations of motion cannot be put into Hamiltonian form. It is only required that some sort of least action principle
be available}..., one can propose a generalization of the Feynman's probability amplitude formula
in the following way:
\begin{equation}\label{FeynmanII}
\mathbf{A}(\mathrm{e}_{_{-}},\,\mathrm{e}_{_{+}})\ \propto\,
\exp{\Bigl\{\frac{i}{\hbar}\,S_{\mathrm{class}}\Bigr\}}\ \boldsymbol{\int} [\mathscr{D}\Sigma]\,\exp{\Bigl\{-\frac{i}{\hbar}\int\limits_{\Sigma}\,\Omega\Bigr\}}\,.
\end{equation}
Here the ``\,surface-summation\,'' is taken over all $\mathscr{CM}$-surfaces $\Sigma$, such that their boundary
contains the classical trajectory $\gamma_{\mathrm{class}}$ and two sideways curves
$\lambda_{_-}$ and $\lambda_{_+}$ within the fibers $\tau^{-1}(\mathrm{e}_{_-})$ and $\tau^{-1}(\mathrm{e}_{_+})$,
respectively.
Using the formula (\ref{Omega}) we can write:
$$
-\int\limits_{\Sigma}\,\Omega\,=\int\limits_{\partial\,\Sigma}\,\Bigl\{\mathbb{T}\,dt+(\partial_{\,\dot{q}^{i}}\,\mathbb{T})\,\alpha^i\Bigr\}
+\int\limits_{\Sigma}\,\mathbb{Q}\,\wedge\,dt\,\equiv
\int\limits_{\partial\,\Sigma}\,\theta_{\,\mathbb{T}}+\int\limits_{\Sigma}\,\mathbb{Q}\,\wedge\,dt
\,.
$$
The first integral term is obviously independent of the choice of the sideways boundary curves $\lambda_{_-}$ and
$\lambda_{_+}$ in $\partial\,\Sigma=\gamma+\lambda_{_+}-\gamma_{\mathrm{class}}-\lambda_{_-}$. Moreover, we can split
the ``\,surface-summation\,'' carried out in (\ref{FeynmanII}) in the following way:
$$
\int[\mathscr{D}\Sigma]\,=\int[\mathscr{D}\gamma]\,\Bigl\{\,\int[\mathscr{D}\Sigma_{\gamma}]\,\Bigr\}\,,
$$
i.e. first we pick out the boundary curve $\gamma$, and then we perform the ``\,summation\,'' over the
subset of those admissible $\mathscr{CM}$-surfaces $\{\Sigma_\gamma\}$ whose world-sheet element $\Sigma_\gamma$
is anchored to the fixed curves $\gamma$ and $\gamma_{\mathrm{class}}$.
After doing this, we get (\ref{FeynmanII}) in the equivalent form:
\begin{equation}\label{FeynmanIII}
\mathbf{A}(\mathrm{e}_{_{-}},\,\mathrm{e}_{_{+}})\ \propto\,
\exp{\Bigl\{\frac{i}{\hbar}\,S_{\mathrm{class}}\Bigr\}}\ \boldsymbol{\int}
[\mathscr{D}\gamma]\,\exp{\biggl\{\frac{i}{\hbar}\Bigl\{\int\limits_{\gamma}-\int\limits_{\gamma_{\mathrm{class}}}\Bigr\}\,\theta_{\,\mathbb{T}}\biggr\}}
\,\boldsymbol{\int}[\mathscr{D}\Sigma_\gamma]\,\exp{\biggl\{\frac{i}{\hbar}\int\limits_{\,\Sigma_\gamma}\,\mathbb{Q}\,\wedge\,dt\biggr\}}\,.
\end{equation}
In the case of conservative forces $\mathbb{Q}\,\wedge\,dt=-d\theta_{\,\mathbb{U}}$,
the surface integral in the last exponent of (\ref{FeynmanIII}) is again only boundary sensitive quantity. Therefore
$$
\mathbf{A}(\mathrm{e}_{_{-}},\,\mathrm{e}_{_{+}})\ \propto\,
\exp{\Bigl\{\frac{i}{\hbar}\,S_{\mathrm{class}}\Bigr\}}\ \boldsymbol{\int}
[\mathscr{D}\gamma]\,\exp{\biggl\{\frac{i}{\hbar}\Bigl\{\int\limits_{\gamma}-\int\limits_{\gamma_{\mathrm{class}}}\Bigr\}\,\bigl(\theta_{\,\mathbb{T}}-\theta_{\,\mathbb{U}}\bigr)\biggr\}}\,\times\,\mathrm{Vol}_{\gamma}
$$
where we have adopted the abbreviated notation
$$
\boldsymbol{\int}\,[\mathscr{D}\Sigma_\gamma]=\mathrm{Vol}_{\gamma}\,=\,\mbox{the
``\,number\,'' of the surfaces containing}\ \gamma-\gamma_{\mathrm{class}}\ \mbox{as the subboundary}\,.
$$
Suppose there are no topological obstructions on the side of $\mathscr{M}$, i.e. that all admissible $\gamma$'s are
homotopically equivalent. Then the factor $\mathrm{Vol}_\gamma$ is $\gamma$-independent, and it might be dropped out
as an infinite constant by normalization. Thus in the case of conservative forces the formula
(\ref{FeynmanIII}) precisely reduces to (\ref{FeynmanI}).

There is still one open point, namely how to express the classical action entering (\ref{FeynmanII}) and (\ref{FeynmanIII})
in terms of $\mathbb{T}$ and
$\mathbb{Q}$. This can be done as follows. Suppose we have the classical trajectory $\gamma_{\mathrm{class}}$ joining
the given events $\mathrm{e}_{_{-}}$ and $\mathrm{e}_{_{+}}$. Let us define
a function assigning to a point $x$ on the trajectory $\gamma_{\mathrm{class}}$ a number $\mathbb{K}\,(x)$:\footnote{Geometrically it is
the integral of the one-form $\mathbb{Q}$ along the curve $\gamma_{\mathrm{class}}$, which is understood as the function
of its upper limit.}
$$
x\,\mapsto\,\mathbb{K}\,(x):=-\int\limits_{\mathrm{e}_{_{-}}}^{\ \ x}\mathbb{Q}\,+\,\mathbb{K}\,(\mathrm{e}_{_-})\,.
$$
The constant $\mathbb{K}\,(\mathrm{e}_{_-})$ might be set to be zero; this plays the same role as the choice of the zero level for the
potential energy in the case of conservative forces. A natural candidate for the classical action then is:
\begin{equation}\label{action}
S_{\mathrm{class}}:=\int\limits_{t_{-}}^{\ \ t_{+}}\,\gamma_{\mathrm{class}}^*\bigl(\mathbb{T}-\mathbb{K}\bigr)\,dt\,.
\end{equation}
We described above all the objects that are necessary for the computation of the probability amplitude. It remains to
give some nontrivial example demonstrating the functionality of (\ref{FeynmanII}) and then
open the discussion.\\
Regarding the example, I have tried to ``\,quantize\,'' the free particle in the one-dimension with the
friction proportional to the actual velocity. To be honest, I finished in a deadlocked when trying to perform
the world-sheet functional integration.
And because I do not have any experience with it, the nontrivial example is unfortunately omitted.
Is there anybody who is able to do it? Please contact me.\\
\hspace{0.9cm}
Let me conclude the paper with few final comments:
\begin{itemize}
\item[$\circ$] the system of co-vectorial annihilators represented by  $\alpha$'s and $\beta$'s (see formulae
               (\ref{alpha-system}) and (\ref{beta-system})) is dependent on the chosen coordinate patch on
               $\mathscr{M}=\mathbb{R}[t]\times\mathcal{Q}$,
               however, the quintessential two-form $\Omega$ is a canonical quantity on $\mathscr{CM}$,
\item[$\circ$] let us consider a function $f: \mathscr{CM}\rightarrow\mathbb{R}$ such that $f$ is non-zero in some
               open subset of $\mathscr{CM}$. The two-forms $\Omega$ and $\Omega^\prime=f\Omega$ have
               common null spaces in this subset and therefore they define equivalent classical dynamics.
               Generally $d\Omega\neq 0$, but in a special case one can succeed in finding an ``\,integrator\,'' $f$
               such that $d\Omega^\prime=0$, i.e. there exists a local one-form $\vartheta$ providing a potential
               for the two-form $\Omega^\prime$.
               The question under what circumstances a Lagrangian function $\mathbb{L}$ exists such that
               $\vartheta=\theta_{\mathbb{L}}$ (an indicator of derivability of dynamics from a variational principle)
               is studied in the inverse problem of calculus of variations,\cite{henneaux1}${}^-$\cite{henneaux2}
\item[$\circ$] the classical trajectory emerging the formula (\ref{I}), and consequently (\ref{FeynmanII}) and
               (\ref{FeynmanIII}), could be replaced by any other fixed curve $\gamma_{\mathrm{ref}}$ within the
               class of admissible curves, however, $\gamma_{\mathrm{class}}$ is privileged by the classical dynamics and
               therefore it is the most natural candidate for the reference point,
\item[$\circ$] here, to be able to talk about the quantum probability amplitudes, one needs to know the solution of
               the classical equations of motion with the given initial condition; in the standard approach, the classical
               solution
               is not necessary for the quantization, it rather
               appears as a saddle point dominating the amplitude in the limit $\hbar\rightarrow 0$,
\item[$\circ$] to see the classical limit in (\ref{FeynmanII}), and an exceptionality of
               the classical history in between the ``\,D-branes\,''
               $\tau^{-1}(\mathrm{e}_{_-})$ and $\tau^{-1}(\mathrm{e}_{_+})$, let us provide an analog of variational principle using
               the two-form $\Omega$:
               \begin{itemize}
               \item[-] consider a class of $\mathscr{CM}$ surfaces $\{\Sigma\}$ such that the boundary
               of any surface in this class is anchored to a chosen reference curve $\gamma_{\mathrm{ref}}$ and
               the D-branes under consideration,\footnote{When substituting $\gamma_{\mathrm{ref}}$ instead of
               $\gamma_{\mathrm{class}}$ on the Figure \ref{figure}, we get the relevant picture for this situation.} i.e.
               $\partial\,\Sigma=\gamma+\lambda_{_+}-\gamma_{\mathrm{ref}}-\lambda_{_-}$,
               \item[-] a stationary surface\footnote{$t$ and $s$ are some worldsheet parameters, here for simplicity we
               identify $t$ with the time coordinate $q^0=t\in\langle t_{_-},t_{_+}\rangle$ over the line element contact
               bundle. A worldsheet ``\,distance\,'' coordinate $s$ is chosen to range within the interval
               $\langle 0,1\rangle$.}
               $\Sigma_{\mathrm{stat}}: (t,s)\mapsto (t=t,\,q^i=q^i(t,s)\,\dot{q}^{\,i}=\dot{q}^{\,i}(t,s))$
               of the action
               $$
               S(\Sigma)=\int\limits_{\Sigma}\,\Omega
               $$
               in the class $\{\Sigma\}$ satisfies:\footnote{We are varying a face of the worldsheet $\Sigma_{\mathrm{stat}}$,
               as well as its boundaries $\gamma$, $\lambda_{_-}$ and $\lambda_{_+}$, since $\Omega|_{\tau^{-1}(\mathrm{e}_{\pm})}=0$
               we get only two equations.}
               \begin{equation}\label{variational eq.}
               \Sigma^{\cdot\,\prime}\,\lrcorner\,d\Omega\,\Bigr|_{\Sigma_{\mathrm{stat}}}=0\,,\ \ \ \ \mbox{and}\ \ \ \ \dot{\gamma}\,\lrcorner\,\Omega=0\,;
               \end{equation}
               here
               $$
               \Sigma^{\cdot\,\prime}=
               \Bigl\{\partial_{t}+
               \frac{\partial q^i}{\partial t}\partial_{q^i}+
               \frac{\partial \dot{q}^{\,i}}{\partial t}\partial_{\dot{q}^{\,i}}
               \Bigr\}\wedge\Bigl\{
               \frac{\partial q^i}{\partial s}\partial_{q^i}+
               \frac{\partial \dot{q}^{\,i}}{\partial s}\partial_{\dot{q}^{\,i}}
               \Bigr\}\ \ \ \mbox{and}\ \ \
               \dot{\gamma}=\Bigl\{\partial_{t}+
               \frac{\partial q^i}{\partial t}\Bigr|_{s=1}\partial_{q^i}+
               \frac{\partial\dot{q}^{\,i}}{\partial t}\Bigr|_{s=1}\partial_{\dot{q}^{\,i}}\Bigr\}
               $$
               stand for a tangent bi-vector to the sought stationary surface $\Sigma_{\mathrm{stat}}$ and
               for a tangent vector to its boundary curve $\gamma$, respectively,
               \item[-] the second equation in (\ref{variational eq.}) is equivalent to (\ref{Lagrange fld.}) and it determines the classical trajectory
               $\gamma_{\mathrm{class}}$,
               \item[-] if we accept in $\{\Sigma\}$ also a degenerate surface (these one is shrunk just to the
               reference curve), then in the special choice of the reference point
               $\gamma_{\mathrm{ref}}=\gamma_{\mathrm{class}}$, we get the solution of
               (\ref{variational eq.}) in the form $\Sigma_{\mathrm{stat}}=\gamma_{\mathrm{class}}$; as a result,
               classical limit, as defined above, is recovered,

               \end{itemize}
\item[$\circ$] composition of the transition amplitudes we are accustomed to from the standard quantum mechanics does
               not work if $d\Omega\neq 0$; this is in accordance with dissipative quantization approaches based on
               various nonlinear generalizations of the Schr\"odinger equation.

%
%\item[$\circ$]
%
%\item[$\circ$]
\end{itemize}

\section*{Acknowledgement}

Many thanks go to Tam\' as F\" ul\" op, Libor \v Sn\' obl, Pavol \v Severa, Peter Pre\v snajder and Vladim\' ir Balek
for their interest, criticism, fruitful discussions and many useful comments. This research was supported in part by
Comenius University Grant UK/359/2006, VEGA Grant 1/3042/06 and M\v SMT \v CR grant LC06002. Special thanks go to
Pavel Exner and Jaroslav Dittrich for their hospitality during my stay at the Nuclear Physics Institute of AS CR.

%\section*{Appendix}
\centerline{$\mathscr{A.\ \ \ \ \ \ \ M.\ \ \ \ \ \ \ D.\ \ \ \ \ \ \ G.}$ }

%\end{multicols}
\end{document}